\documentclass[pre,reprint,showpacs]{revtex4-1}
\usepackage{graphicx}
\usepackage{amsmath,amssymb}
\usepackage{dcolumn}
\usepackage{bm}
\usepackage{hyperref}

\begin{document}

\title{Growth rate distribution of ${\rm NH_4Cl}$ dendrite and its 
scaling structure}

\author{Hiroshi Miki and Haruo Honjo
}
\affiliation{
Department of Applied Science for Electronics and Materials,\\ 
Interdisciplinary Graduate School of Engineering Sciences, \\
Kyushu University, 6-1 Kasuga-Koen, Fukuoka 816-8580, Japan}


\date{\today}
\begin{abstract}
Scaling structure of the growth rate distribution on the interface of a 
dendritic pattern is investigated. The distribution is evaluated for an 
${\rm NH_4Cl}$ quasi-two-dimensional crystal by numerically solving the 
Laplace equation with the boundary condition taking account of the surface 
tension effect. 
It is found that the distribution has multifractality 
and the surface tension effect is almost ineffective in the unscreened large 
growth region. The values of the minimum singular exponent and the fractal 
dimension are smaller than those for the diffusion-limited aggregation pattern. 
The Makarov's theorem, the information dimension equals one, 
and the Turkevich-Scher conjecture between the fractal 
dimension and the minimum singularity exponent hold. 
\end{abstract}


\maketitle

\section{Introduction}
The dendritic pattern is observed in various systems, 
such as crystallization\cite{HG} and electrodeposition\cite{SDG,GBCS}. 
It has been one of the most typical and ubiquitous in nonlinear and 
nonequilibrium physics.  
A dendritic pattern has a stem with the tip growing stably and steadily, 
without splitting. 
Countless sidebranches grow behind the tip due to noise \cite{DKG} and 
instability of a flat interface\cite{MS}. 
These sidebranches compete with the ones around them. 
A longer sidebranch screens off growth of the shorter ones around it.
The competition repeats on various length scale, complicated and hierarchical 
structures are formed \cite{KH} and the pattern becomes fractal as a whole.

The growth process of a dendritic pattern is mainly dominated by diffusion and 
anisotropy. The dendritic pattern is often compared with the diffusion-limited 
aggregation (DLA)\cite{WS} pattern, for which the growth process 
is also dominated by diffusion. However, there is no anisotropy in the DLA 
growth process and the DLA pattern is formed through repeat of tip-splitting. 
It has been reported for an electrodeposition experiment  
that a transition between dendritic pattern and 
DLA pattern is observed as the electrolyte concentration and applied voltage 
are varied\cite{SDG,GBCS}. Furthermore, the dendritic pattern can be formed 
artificially by introducing anisotropy into isotropic viscous fingering
\cite{BJetal,HHK} and, on the other hand, the DLA pattern by removing the 
anisotropy from an anisotropic crystal growth process\cite{HOM}. 

The scaling structure and fractality of dendritic pattern  
have been interesting and important issues for quantitative 
characterization. It has been reported that the area fractal dimension 
$D_f$ of a dendritic pattern with fourfold symmetry is 1.5-1.6 for a 
noise-reduced DLA simulation on a square lattice\cite{Meakin} and 
${\rm NH_4Br}$ crystal growth\cite{Couder_etal}. 
It is clearly less than that of the DLA, $D_f \sim$1.71. 
This is attributed to the fact that the tip is stabilized against splitting 
by anisotropy. Mathematically the lower bound of the fractal dimension for 
the DLA on a square lattice is proved to be 3/2 \cite{Kesten}.  

To characterize a pattern in more detail, the fractal dimension alone is 
insufficient. For the DLA, the growth rate distribution on the interface 
is found to have multifractality in the cases of a simulation\cite{HSM}, 
a crystallization experiment\cite{OH}, and a conformal mapping 
model\cite{BDetal,MHJetal,JMP}. In the present article, we investigate the 
scaling structure of the growth rate distribution, especially compared with 
that of the DLA. We evaluate the growth rate distribution numerically, 
and implement the similar multifractal analysis for a dendritic pattern 
formed in ${\rm NH_4Cl}$ crystallization experiment, 
where the surface tension is effective within a certain length scale. 

\section{Experiment}
\begin{figure}
\includegraphics[width=8cm]{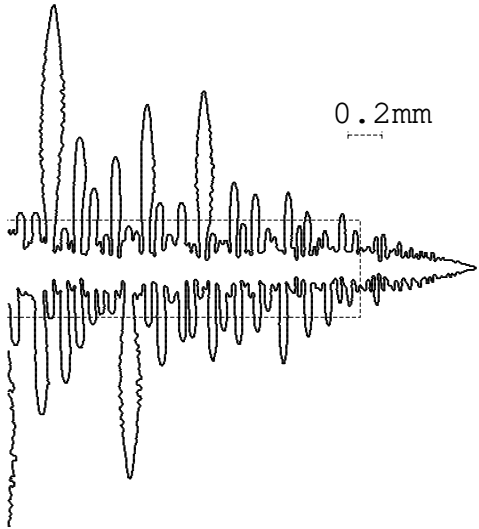}
\caption{Image of a dendritic crystal. The resolution is 
$5.5 \mu {\rm m}$/pixel. Its fractal dimension is 1.54. 
The region inside the broken line is the cutoff region where the surface 
tension effect is neglected in our analysis.
\label{crystal}}
\end{figure}

We use an ${\rm NH_4Cl}$ solution growth dendritic crystal with 
well-developed sidebranches. The details of our experiment are described 
in Ref.\cite{KH}: An ${\rm NH_4Cl}$ aqueous solution saturated at 
approximately $40 { }^\circ {\rm C}$ is sealed, with a nucleation seed left 
in it, in a Hele-Shaw cell, a narrow space between two glass plates. 
The thickness of the cell is 100$\mu {\rm m}$. 
Then when the temperature is lowered (to approximately $30 { }^\circ {\rm C}$), 
the solution becomes supersaturated, nucleation takes place from the seed 
and a growing crystal is observed. The direction of the tip growth is 
$\langle$100$\rangle$ in the supersaturated solution and the growing dendritic 
crystal has fourfold symmetry. Sidebranches grow behind the tip and 
perpendicularly to the stem, with small sub-sidebranches perpendicular to them. 
The image of the crystal is obtained by a microscope and a charge-coupled 
device (CCD) camera, whose resolution is $640 \times 480$ pixels.
The image is binarized by an image processing. The binarized image of a 
crystal interface is shown in Fig.\ref{crystal}. 

The tip growth velocity $v_{\rm tip}$ is $40-49 \mu {\rm m}/{\rm sec}$. 
In the case, sidebranches are well-developed within the shooting window, 
as shown in Fig.\ref{crystal}. 
However since the spacing between them is smaller than the diffusion length, 
they are still competing each other, not growing independently. 
The diffusion length near the tip $l_{\rm D}=2D/v_{\rm tip}$, where $D$ is the 
diffusion constant of ${\rm NH_4Cl}$ ($2.6 \times 10^3 \mu {\rm m}^2/{\rm sec}$
\cite{TanakaSano}) is $\gtrsim 100 \mu {\rm m}$, and that near a sidebranch 
is longer than this. Therefore the growth is regarded as quasi-two-dimensional.

\section{Growth rate evaluation}
In principle it is a faithful method to the original data to evaluate the 
growth rate from the growth site area between two successive images, as 
implemented for a DLA pattern\cite{OH}. However, for a dendritic pattern, 
it is quite difficult to implement with satisfactory precision
due to the limitation of the resolution and since the difference of the growth 
rates between the fast region (around the tips of the stem and longer 
sidebranches) and the slow region (deep inside the forest of sidebranches) is 
much larger than that for the DLA. 
Therefore instead, we evaluate the growth rate $p_{\rm gr}({\bf r}_{\rm int})$ 
at a point ${\bf r}_{\rm int}$ on the interface by the gradient of the 
concentration field $\phi({\bf r})$,
\begin{equation}
p_{\rm gr}({\bf r}_{\rm int}) \sim |\nabla \phi({\bf r})|,
\label{growthrate}
\end{equation}
where $\phi({\bf r})$ is assumed to satisfy the Laplace equation, 
$\nabla^2 \phi({\bf r})=0$, outside of the pattern. This assumption is 
valid if the diffusion length is larger than the characteristic length 
scale of the system, for example, the tip radius of the stem or the average 
spacing of the sidebranch generation. In this case the characteristic length 
scale is of the order of $1\mu {\rm m}$, thus the condition is satisfied.

The Laplace equation for the concentration field is numerically solved by the  
relaxation method on a square lattice, whose spacing is set to be the pixel 
size. The concentration is  supersaturated far away from the interface 
(the saturated concentration at $40 { }^\circ {\rm C}$, 46g per 100g water) 
and the Gibbs-Thomson boundary condition is imposed at the 
interface\cite{Langer,HAK}, 
\begin{eqnarray}
\phi({\bf r}_{\rm int}) &=& \phi_0 (1+d(\theta) \kappa ({\bf r}_{\rm int})),
\label{GT}
\\
d(\theta) &=& d_0 (1-\cos [4\theta]),
\label{stiffness}
\end{eqnarray}
where $\phi_0$ denotes the saturation concentration(at $30 { }^\circ {\rm C}$, 
41g per 100g water), $d(\theta)$ the surface tension coefficient with 
stiffness and fourfold symmetry taken into account, 
$d_0=2.24$ \AA \cite{Dougherty} the capillary length, and 
$\kappa ({\bf r}_{\rm int})$ the local curvature at ${\bf r}_{\rm int}$, 
respectively.
The growth angle $\theta$ is defined as the angle between the growth 
direction at ${\bf r}_{\rm int}$ and that of the stem. 
The curvature and growth angle are calculated by spline interpolation for 
the pixel data of the crystal interface. 

The range in which the growth rate takes value is vast. In the whole 
interface, the ratio of the largest growth rate to the smallest is more 
than $10^{10}$. Even within the region around the tip of the stem, the ratio 
is more than $10^4$. 

When the surface tension effect is taken into account, 
there may be a lattice point where the curvature 
radius is smaller than the lattice spacing. The curvature may vary abruptly 
from point to point around such a lattice point. 
If this situation occurs deep inside the forest of well-developed sidebranches, 
around their roots, a large growth rate, of 
the same order as that of the tips of the stem or longer sidebranches, may be 
generated. This is unnatural since the growth in the region is strongly 
suppressed. In order to appropriately take account of the surface tension 
effect and phenomenologically circumvent the above unnatural situation, 
we consider three cases below:(i) The surface tension is completely 
neglected. In other words, at the interface we set 
$\phi ({\bf r}_{\rm int})=\phi_0$ uniformly. Since the typical length scale of 
the system is of the same order as the lattice spacing and can be regarded as 
the length scale within which the surface tension is effective, this setting 
is reasonable. This case is labelled "Laplace". In this setting the growth 
rate distribution is the harmonic measure. 
(ii) A cutoff $\kappa_c$ is introduced. 
If $|\kappa|>\kappa_c$, $|\kappa|$ is replaced with $\kappa_c$. 
Here we set $\kappa_c=0.01$ times the reciprocal of the lattice spacing. 
This case is labelled "cutoff(1)".
(iii) The surface tension effect is taken into account only around the tips of 
the stem and sidebranches and is neglected deep in the sidebranches, 
in the region shown in Fig.\ref{crystal}. This case is labelled "cutoff(2)". 
The dependence of the results on how to choose $\kappa_c$ for the cutoff(1) 
case and the cutoff region for the cutoff(2) case is very weak
as long as $\kappa_c^{-1}$ is large enough and the cutoff region is wide 
enough, respectively, to suppress the generation of the unnaturally large 
growth rate.
  
\section{Multifractal analysis}
Let us consider that the interface of the dendritic crystal 
is covered by disjoint boxes of size $\epsilon$ and let $p_j(\epsilon)$ 
be the growth rate in the $j$-th box, 
\begin{equation}
p_j(\epsilon) = \sum_{{\bf r}_{\rm int} \in j{\rm -th \quad box}}
p_{\rm gr}({\bf r}_{\rm int}).
\end{equation}
The rate $p_j(\epsilon)$ is normalized to be a probability measure, 
$\sum_j^{N{(\epsilon)}}p_j(\epsilon)=1$, where $N(\epsilon)$ is the number of 
boxes necessary to cover the interface completely. 
The generalized dimension $D(q)$ is defined as\cite{HP} 
\begin{equation}
D(q)=\frac{1}{q-1} \lim_{\epsilon \rightarrow 0} 
\frac{\log Z(\epsilon,q)}{\log \epsilon},
\end{equation}
for $q \ne 1$, where $Z(\epsilon,q)$ is the partition function 
\begin{equation}
Z(\epsilon,q) = \sum_j^{N(\epsilon)} \{p_j(\epsilon)\}^q.
\end{equation}
For $q=1$,
\begin{equation}
D(1) = \lim_{\epsilon \rightarrow 0}
\frac{\sum_j^{N(\epsilon)} p_j(\epsilon)\log p_j(\epsilon)}{\log \epsilon}.
\end{equation}
Practically $D(q)$ is evaluated from the slope of $\log Z(\epsilon,q)$ for 
$\log \epsilon$ by least squares method. 

The singularity exponent $\alpha = \alpha(q)$ and its fractal dimension 
$f(\alpha) = f(\alpha(q))$ are obtained as functions of $q$ by the Legendre 
transformation of the generalized dimension $D(q)$\cite{Halsey_etal}:
\begin{eqnarray}
\alpha(q) &=& \frac{d}{dq}[(q-1)D(q)],
\label{alpha_legendre}
\\
f(\alpha(q)) &=& q\alpha(q)-(q-1)D(q).
\label{f_legendre}
\end{eqnarray}
However, it is not practical to evaluate $\alpha$ and $f(\alpha)$ from 
(\ref{alpha_legendre}) and (\ref{f_legendre}), since it may produce 
relatively large numerical errors. Therefore instead, we adopt a direct 
evaluation method presented in Ref.\cite{CJ} described below. 

First let us construct a new probability measure $\mu_j(\epsilon,q)$ with 
parameter $q$ from $p_j(\epsilon)$ as
\begin{equation}
\mu_j(\epsilon,q) 
= \frac{\{p_j(\epsilon)\}^q}{\sum^{N({\epsilon})}_j \{p_j(\epsilon)\}^q}.
\label{mudef}
\end{equation} 
Then let us define $\zeta(\epsilon,q)$ and $\xi(\epsilon,q)$ as 
\begin{eqnarray}
\zeta(\epsilon,q) &=& \sum_j \mu_j(\epsilon,q) \log[p_j(\epsilon)],
\\
\xi(\epsilon,q) &=& \sum_j \mu_j(\epsilon,q) \log[\mu_j(\epsilon,q)].
\end{eqnarray}
From them $\alpha$ and $f(\alpha)$ are, as functions of $q$, given as 
\begin{eqnarray}
\alpha(q) &=& \lim_{\epsilon \rightarrow 0}
\frac{\zeta(\epsilon,q)}{\log \epsilon},
\label{alpha_direct}
\\
f(q) &=& \lim_{\epsilon \rightarrow 0}
\frac{\xi(\epsilon,q)}{\log \epsilon}.
\label{f_direct}
\end{eqnarray} 
Practically they are evaluated from the slopes of $\zeta(\epsilon,q)$ and 
$\xi(\epsilon,q)$ for $\log \epsilon$, respectively, by least squares method. 
It is easy to show that the definitions (\ref{alpha_direct}) and 
(\ref{f_direct}) satisfy the relations (\ref{alpha_legendre}) and 
(\ref{f_legendre}) by direct calculation.  

\section{Results and discussion}
\begin{figure}
\begin{center}
\includegraphics[width=8.5cm]{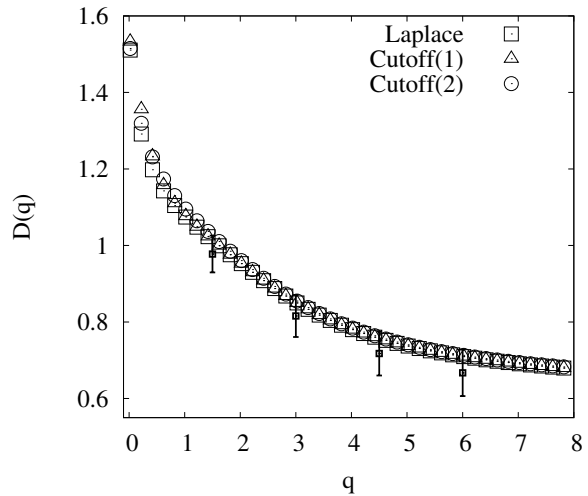}
\caption{Generalized dimension $D(q)$, $q>0$, for the three cases for the 
pattern of FIG.\ref{crystal}.  The increment $\Delta q$ is 0.2.
The error bars are obtained over 13 samples for the "Laplace" case.
\label{gdim}}
\end{center}
\end{figure}

\begin{figure}
\begin{center}
\includegraphics[width=8.5cm]{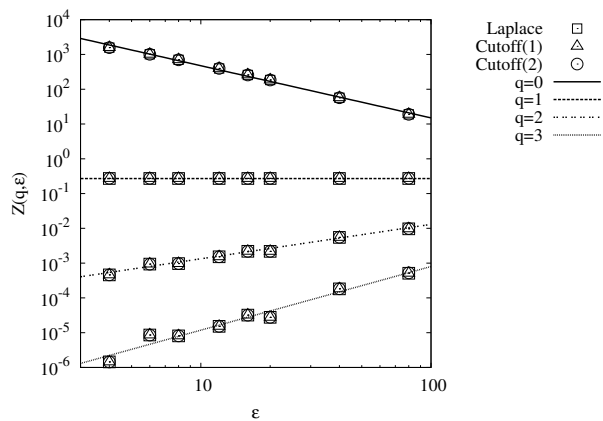}
\caption{Log-log plots of $Z(q,\epsilon)$ for the pattern of FIG.\ref{crystal} 
against the box size $\epsilon$ for $q$=0,1,2, and 3. Note that the slopes 
mean $(q-1)D(q)$. For visibility, the measure is not normalized to be a 
probability ($Z(q=1,\epsilon)$ is the total sum of the growth rate). 
\label{zfit}}
\end{center}
\end{figure}

We are interested in the larger growth rate regime, $q \ge 0$.
For the pattern of Fig.\ref{crystal}, the results of $D(q)$ for the three 
cases are shown in Fig.\ref{gdim}. The log-log plots of $Z(q,\epsilon)$ 
against the box size $\epsilon$ and least squares fitting for several values 
of $q$ are shown in Fig.\ref{zfit}. The box size is chosen from 4 to 80 
pixels, taking the thickness of branches into account. There is a good 
agreement between the results of the three cases.   

\begin{figure}
\begin{center}
\includegraphics[width=8.5cm]{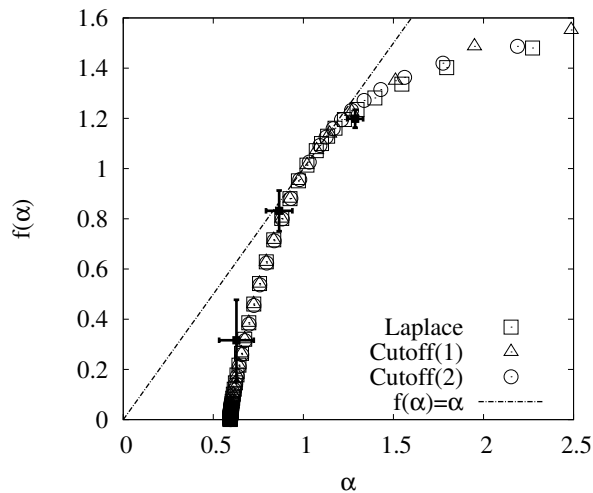}
\caption{Results of the multifractal $f$-$\alpha$ spectrum for the three cases 
for the pattern of FIG.\ref{crystal}.  The contact point with the line 
$f(\alpha)=\alpha$ gives the information dimension 
$D(1)=\alpha(q=1)=f(\alpha(q=1))$. The increment $\Delta q$ is 0.1 for $q<1$ 
and 0.2 for $q>1$. The error bars, corresponding to $q=$0.5, 1.5 and 3, are 
obtained over 13 samples for the "Laplace" case.  
\label{falpha}}
\end{center}
\end{figure}

\begin{figure}
\begin{minipage}{0.8\hsize}
\begin{center}
\includegraphics[width=8cm]{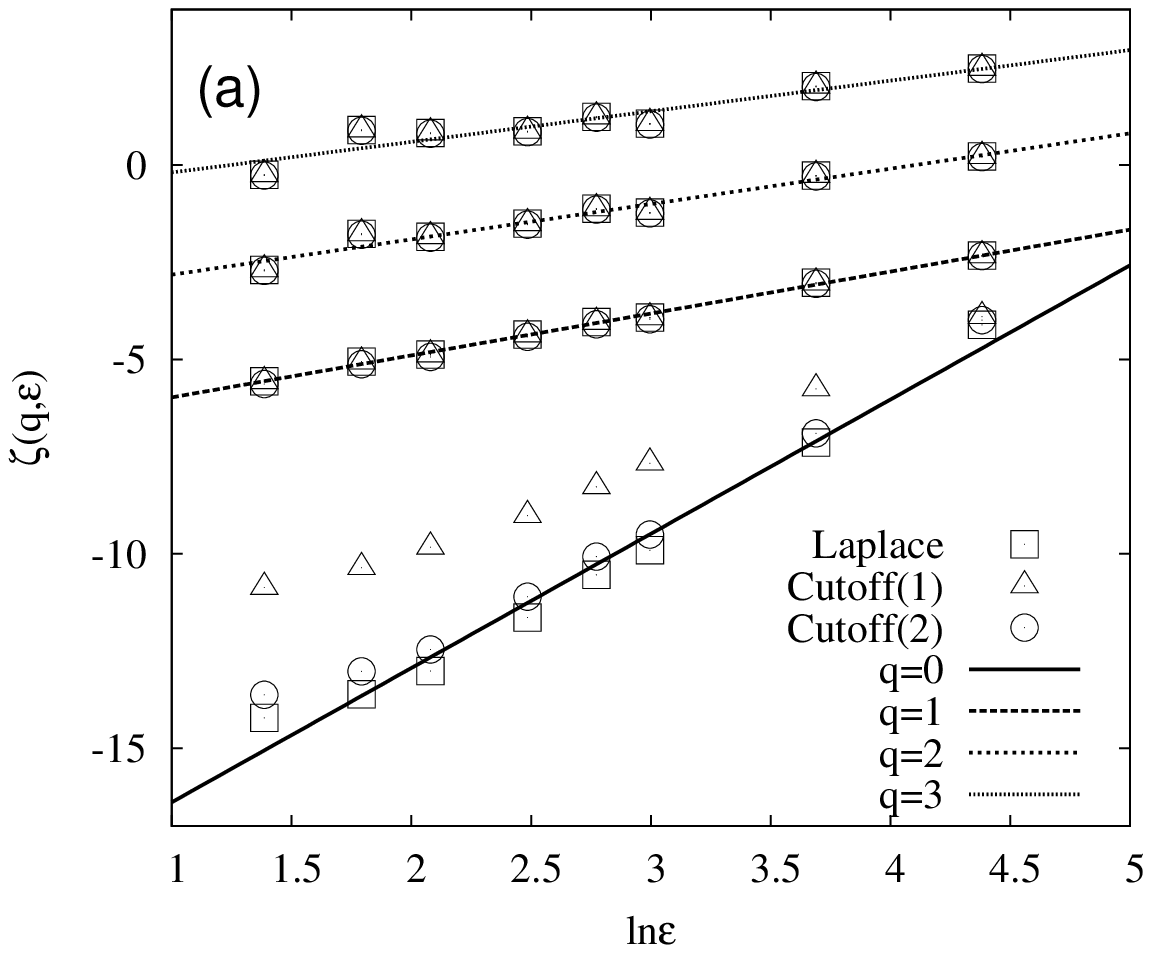}
\end{center}
\end{minipage}
\begin{minipage}{0.8\hsize}
\begin{center}
\includegraphics[width=8cm]{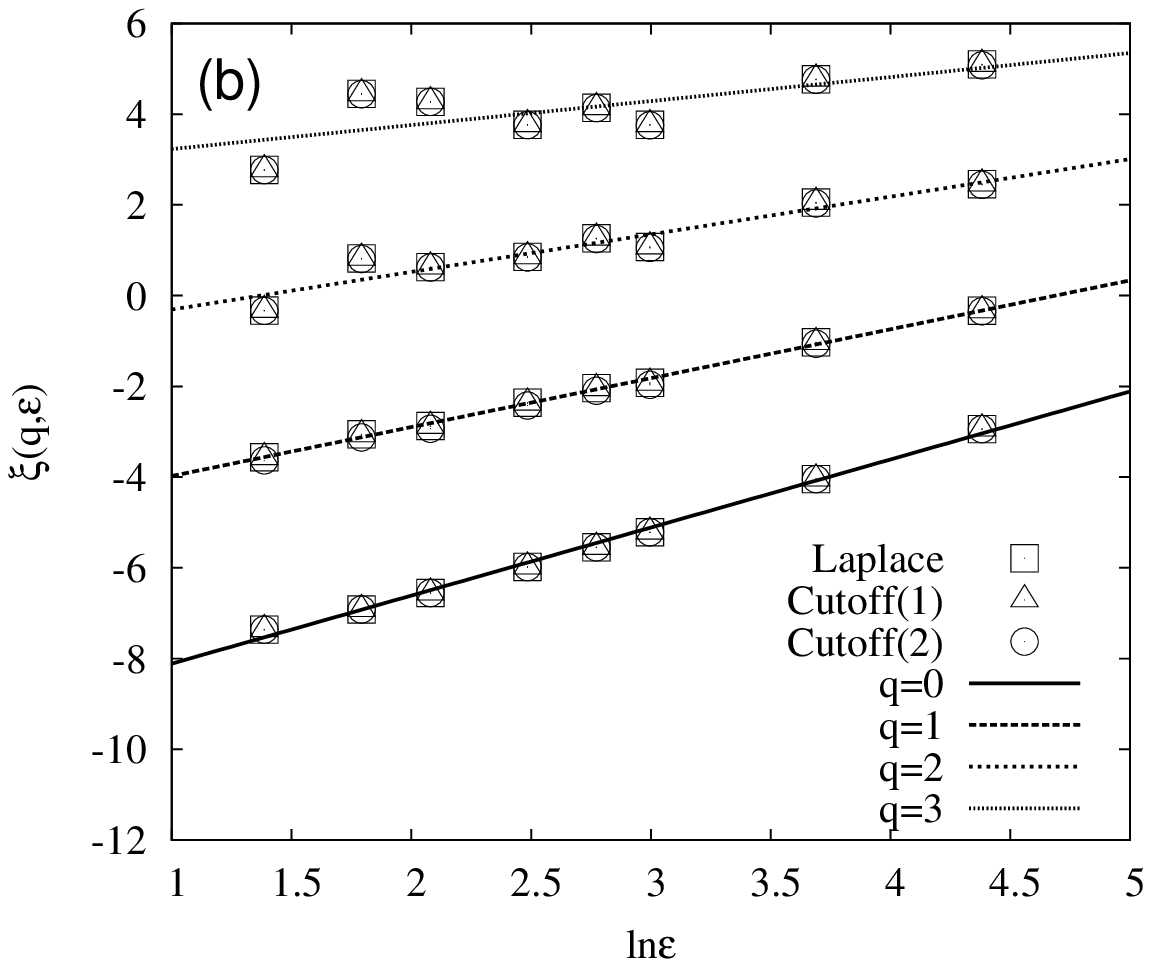}
\end{center}
\end{minipage}
\caption{Fitting of $\alpha$ and $f$ for $q=$0,1,2, and 3. (a)plots of 
$\zeta(q,\epsilon)$ against $\log \epsilon$. (b)plots of $\xi(q,\epsilon)$. 
The fitting lines are shown for the "Laplace" case.
\label{fafits}}
\end{figure}

The results of the multifractal $f$-$\alpha$ spectrum for the three cases for 
the pattern of Fig.\ref{crystal} in the small $\alpha$ region corresponding to 
$q \ge 0$ are shown in Fig.\ref{falpha}. These spectra are evaluated by 
Eqs.(\ref{mudef})-(\ref{f_direct}) and the plots of $\zeta(q,\epsilon)$ and 
$\xi(q,\epsilon)$ against $\log \epsilon$ are shown in FIG.\ref{fafits}. 
There is a good agreement 
between the results of the three cases, except for $\zeta(q=0,\epsilon)$. 
This disagreement is attributed to the surface tension effect, especially the 
contribution of the growth at points deep inside the forest of sidebranches, 
with curvature $|\kappa| \sim \kappa_c$. 
The relatively small difference of $\zeta(q=0,\epsilon)$ between for the 
"Laplace" and the "Cutoff(2)" case and the agreement of $\zeta(q,\epsilon)$, 
q=1,2, and 3, for the three cases indicate that the surface tension effect is 
almost ineffective in the unscreened large growth region. The fact that $D(q)$ 
and $f(\alpha)$ take continuous values and depend on $q$ or $\alpha$ means that 
the growth rate distribution has multifractality.    

Some characteristic values for the scaling exponents are summarized in 
Table.\ref{table1}, along with those for DLA conformal mapping model
\cite{MHJetal,JMP} for comparison. 
Both the fractal dimensions of area $D_f$ and perimeter length $D(0)$ 
are about 1.5, manifestly smaller than those for the DLA. 
This result agrees with the results of on-lattice simulation\cite{Meakin} 
and mathematics\cite{Kesten}. 
\begin{table}
\begin{tabular}{ccccc}
Case&$D_f$&$D(0)$&$D(1)$&$\alpha_{\rm min}$\\
\hline
Laplace&1.55$\pm$0.04&1.53$\pm$0.02&1.02$\pm$0.03&0.56$\pm$0.04\\
Cutoff(1)&1.55$\pm$0.04&1.53$\pm$0.02&1.03$\pm$0.03&0.56$\pm$0.04\\
Cutoff(2)&1.55$\pm$0.04&1.53$\pm$0.02&1.07$\pm$0.03&0.58$\pm$0.05\\
DLA&1.713$\pm$0.003&$\sim$1.71&-&0.665$\pm$0.004
\end{tabular}
\caption{List of characteristic scaling exponents. 
The averages and errors are obtained over 13 samples.
Note that $D_f$ is the fractal dimension 
of the area of the pattern and $D(0)$ is the fractal dimension of the 
interface, on which the growth rate measure is defined, 
and that by definition, $\alpha_{\rm min}=D(q \rightarrow \infty)$. 
The values for DLA are cited from the results of a 
conformal mapping model\cite{MHJetal,JMP}.
\label{table1}}
\end{table} 
The smallest singularity exponent is obtained at the tip of the stem where 
the growth is most active and it is also smaller than that for the DLA. 
Furthermore, this agrees with the Turkevich-Scher scaling conjecture 
$D_f=1+\alpha_{\rm min}$\cite{TS,HMP},
which argues that the fractal dimension depends only on the scaling 
behavior of the growth of the most active domain. Note that it is clear that 
the tip of the stem is the most actively growing domain for a dendritic 
pattern, while it is reported that for the DLA, the most active growth domain 
is not the outermost tip\cite{JMP}. 
The information dimension $D(1)$ is regarded as the fractal dimension of 
the active zone where the growth is not screened\cite{CS}. It is  proved 
by Makarov\cite{Makarov} that exactly $D(1)=1$ for the harmonic measure. 
Our results agree with the theorem, remarkably even though the surface 
tension is taken into account.

\section{Conclusion}
We evaluated the growth rate distribution of an ${\rm NH_4Cl}$ dendritic 
crystal interface by numerically solving the Laplace equation and investigated 
its scaling property. The effect of the surface tension is taken into account 
as the Gibbs-Thomson boundary condition with some types of cutoff introduced 
based on phenomenologically plausible assumptions. We found that in the 
unscreened large growth rate regime the distribution has 
multifractality and the surface tension effect is not essential. 
The fractal dimension and the value of the smallest singular
exponent are smaller than that of the DLA and consistent 
with the previous results given in theory and simulation. Our results agree 
with the Makarov's theorem for the harmonic measure, $D(1)=1$, and the 
Turkevich-Scher scaling conjecture, $D_f=1+\alpha_{\rm min}$ in spite of the 
surface tension effect.  

\begin{acknowledgments}
This research was supported by the Japan Ministry of Education, Culture, 
Sports, Science and Technology, Grant-in-Aid for Scientific Research, 
No. 21540392. 
\end{acknowledgments}


\end{document}